\documentclass[12pt, a4paper]{article}
\usepackage[utf8]{inputenc}
\usepackage{fancyhdr}
\usepackage[english]{babel}
\usepackage[table]{xcolor}
\usepackage{hyperref} 
\usepackage{enumerate}
\usepackage{enumitem}
\usepackage{blkarray}
\usepackage[bf]{caption}
\usepackage{url}
\usepackage{graphicx}
\usepackage{lipsum}
\usepackage{amsmath}
\usepackage{mathtools}
\usepackage{tabularx}
\usepackage{amsfonts}
\usepackage[paper=a4paper, left=25mm, right=22mm, top=25mm, bottom=25mm]{geometry}
\usepackage{amssymb}
\usepackage{natbib}
\usepackage[bottom]{footmisc}
\usepackage{subfig}
\usepackage{multirow}

\newcommand{\betab}{{\pmb{\beta}}}

\newcommand{\lambdab}{\pmb{\lambda}}

\newcommand{\varthetab}{\pmb{\mathit\vartheta}}

\newcommand{\yb}{\textbf{y}}

\newcommand{\blanco}[1]{}

\newcommand\norm[1]{\left\lVert#1\right\rVert}

\author{Hendrik van der Wurp}

\title{Generalised joint regression for count data with a focus on modelling football matches}

\begin{document}
\captionsetup{format=hang}
\renewcommand{\baselinestretch}{1}\normalsize
\vspace{1cm}

\noindent{\LARGE\textbf{Generalised Joint Regression for Count Data with a Focus on Modelling Football Matches}}\\

\noindent
\textbf{{\large Hendrik van der Wurp\textsuperscript{1}, Andreas Groll\textsuperscript{1}, Thomas Kneib\textsuperscript{2}, \linebreak Giampiero Marra\textsuperscript{3} and Rosalba Radice\textsuperscript{4}}} \\ \vspace{0.5cm}

\noindent
\textsuperscript{1} Faculty of Statistics, TU Dortmund University\\
\textsuperscript{2} Chair of Statistics, Georg-August University G\"ottingen\\
\textsuperscript{3} Department of Statistical Science, University College London\\
\textsuperscript{4} Faculty of Actuarial Science and Insurance, Cass Business School, City University of London
\vspace{0.75cm}
\hrule
\vspace{0.25cm}
\noindent
\textbf{Address for correspondence:} Hendrik van der Wurp, Faculty of Statistics, TU Dortmund University, Vogelpothsweg 87, 44227 Dortmund, Germany.\\
\noindent
\textbf{E-mail:} vanderwurp@statistik.tu-dortmund.de.\\
\textbf{Phone:} (+49) 231 755 7204.\\
\textbf{Fax:} (+49) 231 755 5305.

\vspace{0.75cm}
\hrule
\vspace{0.25cm}
\noindent
\textbf{Abstract:}
We propose a versatile joint regression framework for count responses. The method is implemented in the \texttt{R} add-on package \texttt{GJRM} and allows for modelling linear and non-linear dependence through the use of several copulae. Moreover, the parameters of the marginal distributions of the count responses and of the copula can be specified as flexible functions of covariates. Motivated by a football application, we also discuss an extension which forces the regression coefficients of the marginal (linear) predictors to be equal via a suitable penalisation. Model fitting is based on a trust region algorithm which estimates simultaneously all the parameters of the joint models. We investigate the proposal's empirical performance in two simulation studies, the first one designed for arbitrary count data, the other one reflecting football-specific settings. Finally, the method is applied to FIFA World Cup data, showing its competitiveness to the standard approach with regard to predictive performance.\medskip
\hrule

\vspace{0.5cm}


\section{Introduction}

There are many data situations where bivariate (or even multivariate) counts are the end point of interest and a priori assuming independence between such variables may be questionable. In particular, in many team sports such as football, handball or ice
hockey, one usually jointly observes the number of goals (or, more generally, the number of points as, for instance, in basket ball or American football) of both competing teams. These are certainly associated as the final scores are the outcome of many single game situations where the players of both teams are involved in. For example, in football it might be realistic to assume that if one team scores and takes the lead then the other team increases its effort to also score in order not to lose the game.

Historically, in modelling football scores Poisson distribution approaches are well established 
and have been widely used, see e.g., \citet{Lee:97} or \citet{Dyte:2000}, who 
modelled the number of the teams' goals with independent Poisson distributions. 
\citet{DixCol:97} were among the first to investigate dependency between scores of competing teams.
They expanded the independent Poisson approach by an additional dependence parameter
to adjust for certain under- and overrepresented match results.

In this article, we present a flexible generalised joint regression framework for count responses. The linear or non-linear dependence between the outcomes is modelled via means of copulae. Moreover, all the parameters of the joint distribution can be specified as flexible functions of covariates. Motivated by our case study, we also provide an extension of the method which enforces the linear regression coefficients of the marginal predictors to be equal by introducing a penalty on the pairwise differences of the relevant effects. This is indeed particularly useful for modelling team sports data where the predictors of both competing teams are usually based on the same set of covariates whose effects are often assumed to be equal \citep[e.g.,][]{GrollEtAl2018}. The proposed method is incorporated in the \texttt{R} package \texttt{GJRM} \citep[Generalized Joint Regression Modelling,][]{Mar:2017:gjrm}. 

A first approach for explicitly accounting for dependencies in football using the bivariate Poisson distribution was proposed by \citet{KarNtz:2003}. \citet{mchale2007modelling} employed instead copula models with Poisson margins to model shots-for and shots-against. Copula models have already been considered in the context of count responses (see, for example, \cite{NK2010} and \cite{trivedi2017note} and references therein). Here, we elected to extend the modelling and computational framework of \cite{MaRa:2017} to the case of discrete margins because it allows for several types of covariate effects, for a rich variety of copula functions, and for any type of quadratic penalty which was a crucial aspect in our case study. 

We investigate the proposed method's empirical performance in two simulation studies, the first one which does not assume equal regression coefficients and the other one which does instead, hence reflecting football-specific settings. Finally, the method is applied to FIFA World Cup data which shows that assuming equal coefficients yields a superior predictive performance when compared to the approach that does not impose such equality. Bookmakers are used as a natural benchmark for our model, and profits derived from adopting a given betting strategy are calculated.

The rest of the manuscript is structured as follows. The methodological background of the joint regression framework for count responses and the penalty extension specifically designed for football data are introduced in Section~\ref{sec:methods}. In Section~\ref{sec:simulation}, we present two simulation studies which analyse the proposed method's predictive performance in different data settings. The data employed in our football application, covering all matches of the five preceding FIFA World Cups 2002 -- 2018, are described in Section~\ref{sec:data}. Using these data, we then compare the predictive performance of the copula models whose the parameters of the marginal distributions are assumed to be equal and not (Section~\ref{sec:pred:perf}). Assuming equal regression coefficients yields a superior performance, as elaborated in Section~\ref{sec:appl:results}. We conclude in Section~\ref{sec:conclusion}.

\section{Methodology}\label{sec:methods}

This section provides the most salient details of the proposed generalised joint regression modelling framework for count data. In particular, motivated by our case study, which uses FIFA World Cup football data, we will focus on the methodological aspects that are relevant to the model specification adopted in Section \ref{sec:application}. Note that we have considered single-parameter copulae and marginal distributions with up to two parameters since they were deemed appropriate for our empirical application. However, the computational framework can be conceptually easily extended to copulae and marginal distributions with more parameters.


\subsection{Model structure and estimation approach}\label{sec:basic:model}

For notational convenience, we drop the conditioning on parameters (of the marginal distributions and of the copula function) and
observation index $i$. It is clear, however, from the context of the paper that bivariate count data with integer realisations $\yb_i=(y_{i1}, y_{i2})^T, i = 1,\ldots, n$, for a sample of size $n$, are available for modelling purposes and that covariate effects have to be accounted for. 

We assume that the joint cumulative distribution function (cdf) $F(\cdot,\cdot)$ of two discrete outcome variables, $Y_1 \in \mathbb{N}_0$ and $Y_2 \in \mathbb{N}_0$ can be expressed as
\[
P(Y_1 \leq y_1, \ Y_2 \leq y_2) \ = \ C_\theta \left( P(Y_1 \leq y_1), \ P(Y_2 \leq y_2) \right) \ = \ C_\theta(F_1(y_1), \ F_2(y_2))\,,
\]
where $F_1(\cdot)$ and $F_2(\cdot)$ are the marginal cdfs of $Y_1$ and $Y_2$ taking values in $(0, 1)$, $C_\theta: (0,1)^2 \rightarrow (0,1)$ is a two-place copula function which does not depend on the marginals, and $\theta$ denotes the copula parameter measuring the dependence between the two random variables. The adopted dependence structure relies on $C_\theta(\cdot,\cdot)$ and its parameter $\theta$; the copulae implemented in \texttt{GJRM} are reported, for instance, in Table 1 of \cite{MRjasa}. It should be pointed out that in a setting with discrete marginal distributions the copula function $C_\theta$ is not unique (see \citealp{SchwSklar:1983}; Chapter 6). However, as elaborated by several authors including \cite{NK2010} and \cite{trivedi2017note}, this is not an issue of practical concern in regression settings. Following \citet{trivedi2017note}, the joint probability mass function (pmf) $c_\theta(\cdot,\cdot)$ for a given copula $C_\theta$ on the two-dimensional integer grid can be obtained as
\begin{align}\label{Copequ}
c_\theta(F_1(y_1), \ F_2(y_2)) = & & & C_\theta(F_1(y_1), \ F_2(y_2)) - C_\theta(F_1(y_1-1), \ F_2(y_2)) \nonumber \\
&&& - C_\theta(F_1(y_1), \ F_2(y_2-1)) + C_\theta(F_1(y_1-1), \ F_2(y_2-1)).
\end{align}

For the outcome variables $Y_1$ and $Y_2$, we have considered (and implemented in \texttt{GJRM}) four main discrete distributions, namely Poisson, negative binomial type I, negative binomial type II and Poisson inverse Gaussian; these have been parametrised according to \cite{RS2005}. In the following, we focus on Poisson marginals since they were found to be appropriate for modelling our count responses (see Section \ref{sec:application}).

Let now the parameters of the two marginal distributions as well as of the copula parameter $\theta$ be connected with sets of covariates of sizes $p_1, p_2$ and $p_{\theta}$, respectively. Moreover, let the corresponding covariate vectors be denoted by $\boldsymbol{x}_1$, $\boldsymbol{x}_2$ and $\boldsymbol{x}_\theta$, including entries for intercepts and/or dummy variables for categorical variables. For two Poisson-distributed margins with rate parameters $\lambda_1$ and $\lambda_2$ and a copula function characterised by one parameter, we may have 
\begin{align}\label{eq:lin:mod}
\log(\lambda_1) & = \eta_1  = \beta_0^{(1)} + x_{1}^{(1)} \beta_1^{(1)} + \ldots + x_{p_1}^{(1)} \beta_{p_1}^{(1)} =  (\boldsymbol{x}^{(1)})^{T}\boldsymbol{\beta}^{(1)}\,, \nonumber\\
\log(\lambda_2) & = \eta_2 =  \beta_0^{(2)} + x_{1}^{(2)} \beta_1^{(2)} + \ldots + x_{p_2}^{(2)} \beta_{p_2}^{(2)} =  (\boldsymbol{x}^{(2)})^{T}\boldsymbol{\beta}^{(2)}\,, \\
g(\theta) & = \eta_\theta = \beta_0^{(\theta)} + x_{1}^{(\theta)} \beta_1^{(\theta)} + \ldots + x_{p_\theta}^{(\theta)} \beta_{p_\theta}^{(\theta)} =  (\boldsymbol{x}^{(\theta)})^{T} \boldsymbol{\beta}^{(\theta)}\,,\nonumber
\end{align}
where $\boldsymbol{\beta}^{(1)}, \boldsymbol{\beta}^{(2)}$ and $\boldsymbol{\beta}^{(\theta)}$
are $p_1$-, $p_2$- and $p_{\theta}$-dimensional vectors of regression effects, respectively.
The logarithmic link function guarantees positivity of the two Poisson parameters $\lambda_1$ and $\lambda_2$.
Other distributions may require different link functions. The vectors $\boldsymbol{x}^{(1)}$,
$\boldsymbol{x}^{(2)}$ and $\boldsymbol{x}^{(\theta)}$ are subsets of a complete set of covariates
$\boldsymbol{x}$ of size $d$, with $p_1+p_2+p_\theta = k \geq d$. Finally, $g(\cdot)$ is a link function whose choice will depend on the employed copula (see \citealp{MRjasa}).

We would like to stress that the equations in \eqref{eq:lin:mod} represent a substantial simplification of the possibilities allowed for in the proposed modelling framework. In particular, our implementation allows to include non-linear functions of continuous covariates, smooth interactions between continuous and/or discrete variables and spatial effects, to name but a few. For this purpose, the penalised regression spline approach was adopted and the reader is referred to, e.g., \cite{MaRa:2017} for some examples. Due to the specific type of penalisation employed in this paper (see the next section), in this work we focus on linear effects as presented in \eqref{eq:lin:mod}.

The model's log-likelihood for the $k$-dimensional vector $\boldsymbol{\beta}^T=\left((\boldsymbol{\beta}^{(1)})^T, (\boldsymbol{\beta}^{(2)})^T, (\boldsymbol{\beta}^{(\theta)})^T\right)$ is
\begin{equation}\label{eq:loglik}
\ell(\boldsymbol{\beta}) = \sum_{i=1}^n \log \left\{ c_\theta\left( F_1(y_{i1}), F_2(y_{i2})\right)\right\}\,,
\end{equation}
where, for $j = 1,2$, 
\[
F_j(y_{ij}) = \exp(- \exp( \eta_{ij}) ) \sum_{m=0}^{y_{ij}} \dfrac{\exp( \eta_{ij})^m}{m!}\,.
\]
If spline terms appear in the model specification then (\ref{eq:loglik}) has to be augmented by a quadratic penalty term whose role would be to enforce specific properties on the respective functions, such as smoothness.

Simultaneous estimation of all the parameters is based on maximising $\ell(\boldsymbol{\beta})$ 
with respect to $\boldsymbol{\beta}$. To this end, we extended the estimation approach of
 the \texttt{R} package \texttt{GJRM} \citep{Mar:2017:gjrm} to accommodate discrete margins. 
 The fitting algorithm is based on iterative calls of a trust region algorithm, which requires 
 first and second order analytical derivatives, which have been tediously derived and verified 
 numerically. In \texttt{R}, the algorithm is realised in the \texttt{trust()} function from the \texttt{trust} package by \citet{trust}. The modularity of the implementation means that, in principle, it will be easy to extend our modelling framework to parametric copulae and discrete marginal distributions not included in the package. To facilitate the computational developments, when evaluating \eqref{Copequ}, we replaced $F_j(y_j-1)$ with $F_j(y_j) - f_j(y_j)$ for $j=1,2$, where $f_j(\cdot)$ denotes the $j^{th}$ marginal pmf. This is especially relevant for the case $y_j = 0$ where $F_j(-1)$ needs to be set to 0. 

As hinted above, the \texttt{GJRM} infrastructure allows one to incorporate any quadratic penalty of the form $\frac{1}{2}\boldsymbol{\beta}^{T}\boldsymbol{S}\boldsymbol{\beta}\,$, where $\boldsymbol{S}$ is a penalty matrix. The next section discusses a specification of penalty which is particularly useful for modelling football matches.

\subsection{A penalty approach for football data}\label{sec:penalty}

The model adopted in Section \ref{sec:application} is based on $F(y_1,y_2|\varthetab) = C\left(F_1(y_1|\lambda_1), F_2(y_2|\lambda_2);\theta\right)\,$, where $Y_1 \sim \text{Poi}(\lambda_1)$, $Y_2 \sim \text{Poi}(\lambda_2)$ and $\varthetab = (\lambda_1, \lambda_2, \theta)^T$. Each Poisson marginal models the number of goals
scored by team $j\in\{1,2\}$ and is characterised by parameter $\lambda_j$. The expected number of goals for team $j$ in a match $i$ is given by
\[
 \lambda_{ij}  = \exp \left(\beta_0^{(j)} + x_{i1} \beta_1^{(j)} + \ldots + x_{ip} \beta_p^{(j)} \right), \qquad i=1,\ldots,n, \qquad j=1,2\,.
 \]
Although including covariate information into $\theta$ is possible and allowed for by \texttt{GJRM}, for simplicity, the copula parameter $\theta$ is specified as function of an intercept $\beta_0^{(\theta)}$ only. This way, we achieve explicit comparability of dependence strengths in terms of Kendall's $\tau$ among different copula functions. 

In contrast to the setting of the equations in \eqref{eq:lin:mod}, in football it is sensible to consider the same set of covariates for both competing teams (i.e., $p_1=p_2=:p$). Also, one needs to impose the same covariate effects across the predictors of the marginal distributions. Specifically, assuming covariates that are ordered such that $x_{ir}^{(1)}$ and $x_{ir}^{(2)}, r=1,\ldots,p,$ correspond to the same
regressors (e.g., the average age of team 1 and team 2, respectively), we would like to achieve $\beta_r^{(1)}=\beta_r^{(2)}$ for all
$r \in 0,\ldots,p$. Without this restriction, being first- or second-named team could affect the estimation of $\beta_k^{(j)}$ and thus make the interpretation of the coefficients questionable, as stressed in \citet{GrollEtAl2018}.

Equal (or virtually equal) coefficients for both margins can be achieved using a properly defined penalty. To this end, we propose to use the following penalised version of log-likelihood \eqref{eq:loglik}, i.e.
\begin{equation}\label{eq:lasso}
\ell_p(\boldsymbol{\beta}) = \ell(\boldsymbol{\beta}) - \frac{1}{2}\, \xi\, \sum_{j=0}^{p} w_j\left(\beta_j^{(1)} - \beta_j^{(2)}\right)^2\,,
\end{equation}
where the ridge-type penalty acts on the differences of the pair of intercepts as well as the pairs of coefficients corresponding to the same covariates, with suitably chosen weights $w_j$ and penalty parameter $\xi$. This penalty can be easily incorporated in \texttt{GJRM} via a suitably designed penalty matrix $\boldsymbol{S}$, which in this case is equal to
\begin{equation}\label{eq:pen:matrix}
\boldsymbol{S} = \xi \cdot \boldsymbol{W} \circ\,\begin{blockarray}{ccccccccc}
\beta_0^{(1)} & \beta_1^{(1)} & \ldots & \beta_p^{(1)} & \beta_0^{(2)} & \beta_1^{(2)} & \ldots &\beta_p^{(2)} & \beta_0^{(\theta)} \\
 \begin{block}{(cccc|cccc|c)}
1 & 0  & \ldots & 0 & -1 & 0 & \ldots & 0 & 0 \\
0 & 1 &  \ldots & 0 & 0 & -1 & \ldots & 0 & 0 \\
\vdots & \ddots  & \ddots & \ddots & \ddots & \ddots & \ddots & \vdots & \vdots\\
0 & 0  & \ldots & 1 & 0 & 0 & \ldots & -1 & 0\\ \BAhline
-1& 0 & \ldots & 0 & 1 & 0 & \ldots & 0 & 0 \\
0 & -1 & \ldots & 0 & 0 & 1 & \ldots & 0 & 0 \\
\vdots & \ddots & \ddots & \ddots & \ddots & \ddots & \ddots & \vdots & \vdots\\
0 & 0  & \ldots & -1 & 0 & 0 & \ldots & 1& 0\\ \BAhline
0 & 0 & 0 & 0 & 0 & 0 & 0 & 0 & 0 \\
\end{block}
\end{blockarray}\,\, ,
\end{equation}
where ``$\circ$" denotes the Hadamard matrix product. Moreover, $\xi$ is a tuning parameter controlling
the strength of the penalty, and $\boldsymbol{W}$ a weight matrix of the form
\[ 
\boldsymbol{W} = \begin{pmatrix}
\boldsymbol{w}^T&  \boldsymbol{w}^T & 0 \\
\vdots & \vdots &  \vdots \\
\boldsymbol{w}^T &  \boldsymbol{w}^T & 0 \\
0 & \ldots  & 0
\end{pmatrix} \,,
\]
which is of the same dimension as the other matrix from the Hadamard product in \eqref{eq:pen:matrix}.
The vectors of weights $\boldsymbol{w} \in \mathbb{R}^{p+1}$ 
depend on the current fit $\hat\betab^{[k]}$, which is obtained at iteration $k$ of the algorithm.
 In order to shrink all paired differences jointly to zero, we use 
$w_j=\left|\hat\beta_j^{(1)} - \hat\beta_j^{(2)}\right|$ (here suppressing the iteration index 
for notational convenience). If the penalty parameter $\xi$ is chosen sufficiently large, 
we obtain (virtually) equal parameter estimates. In our case study $\xi$ was set to $10^9$. 


\subsection{Prediction}\label{sec:predict}


After fitting a model with equal or unequal coefficients, we can calculate probabilities for each possible pair of outcomes. We will sketch this modus operandi for our football setting, but it could be easily generalised to different data situations and marginal distributions.
First, based on the two teams' estimated coefficients $\hat{\boldsymbol{\beta}}^{(j)}, j=1,2,$
for an arbitrary match $i$, we estimate the marginal Poisson parameters $\lambda_1$ and $\lambda_2$ using 
\[ \widehat{\lambda}_{i1} = \exp(\widehat{\eta}_i^{(1)}) = \exp\left((\boldsymbol{x}_{i}^{(1)})^{T}\hat{\boldsymbol{\beta}}^{(1)}\right)\,, \]
\[ \widehat{\lambda}_{i2} = \exp(\widehat{\eta}_i^{(2)}) = \exp\left((\boldsymbol{x}_{i}^{(2)})^{T}\hat{\boldsymbol{\beta}}^{(2)}\right)\,.\]

\noindent We then use the \texttt{copula} package \citep{R:copula} to obtain the joint function for a specific chosen copula with Poisson margins and parameters $\widehat{\lambda}_{i1}, \widehat{\lambda}_{i2}$ and $\widehat{\theta}$. The probability for a specific match outcome $(y_1,y_2)$ can be calculated using the joint pmf described in Section~\ref{sec:basic:model}. That is,
\begin{align}\label{eq:copula:pred}
P\left((Y_1,Y_2)=(y_1,y_2)\right) & =   c(F_1(y_1),F_2(y_2);\widehat{\theta}) \nonumber\\
& = C(F_1(y_1),F_2(y_2);\widehat{\theta}) - C(F_1(y_1-1),F_2(y_2);\widehat{\theta}) \nonumber\\
& - C(F_1(y_1),F_2(y_2-1);\widehat{\theta}) + C(F_1(y_1-1),F_2(y_2-1);\widehat{\theta})\,,
\end{align}
where $F_1(\cdot)$ and $F_2(\cdot)$ are the Poisson cdfs with corresponding parameters $\widehat{\lambda}_{i1}$ and $\widehat{\lambda}_{i2}$. 

In football it is typically also of high interest to obtain estimates of the three coarser match results \textit{win}, \textit{draw}
and \textit{defeat} (from the perspective of the first-named team); these are collected in the categorical (ordinal) outcomes $\tilde{y}_i \in \{1,2,3\}$.
For each of these categories, the probabilities of all relevant precise match results $(y_1,y_2)$ are simply added up. Because the calculation of the pmf for all possible results may quickly become computationally infeasible, for practical purposes, in the football application of Section \ref{sec:pred:perf}, we only consider $y_1,y_2\leq 20$. The results $\widehat{\pi}_{i1}$, $\widehat{\pi}_{i2}$ and $\widehat{\pi}_{i3}$ are estimates for the true probabilities
\begin{align*}
P(\textit{win}) & = P(\tilde{Y}_i = 1) = \pi_{i1}\,, \\
P(\textit{draw}) & = P(\tilde{Y}_i = 2) = \pi_{i2}\,,  \\
P(\textit{defeat}) & = P(\tilde{Y}_i = 3) = \pi_{i3}\,,
\end{align*}
which can then be used, e.g., for comparison with bookmakers' odds.

\vspace{0.5cm}

\noindent Appendix \ref{software} provides a short operational description of the software.

\section{Simulation Study}\label{sec:simulation}

In this section, we present two sets of simulations for pairs of (marginally) Poisson-\linebreak distributed outcomes. The first one shows that the proposed estimation approach is able to (a) select the correct copula and (b) deliver estimates that are close to the true coefficients $\boldsymbol{\beta}$. The second set shows that when the true underlying coefficients are equal across margins, a penalisation, as presented in Section~\ref{sec:penalty}, improves considerably goodness-of-fit.

For both simulations, we draw covariates $x_1, \ldots, x_6$ from independent uniform distributions on the interval
$[0,1]$, i.e.\ $x_{r}\sim \mathcal{U}[0,1]$, for $n=250$ observations and $r=1,\ldots 6$. For each observation, the
Poisson parameters $\lambda_{ij}$ (${i=1,\ldots,250}$, $j=1,2$) are determined with given coefficient vectors
${\boldsymbol{\beta}^{(1)} = (\beta_0^{(1)}, \beta_1^{(1)}, \beta_2^{(1)}, \beta_3^{(1)})^T}$
and ${\boldsymbol{\beta}^{(2)} = (\beta_0^{(2)}, \beta_1^{(2)}, \beta_2^{(2)}, \beta_3^{(2)})^T}$ via
\begin{align}\label{eq:sim:pred1}
 \lambda_{i1} & = \exp\left(\beta^{(1)}_0 + x_{i1}\beta_1^{(1)} + x_{i2}\beta_2^{(1)} + x_{i3}\beta_3^{(1)}\right)\,, \\
 \lambda_{i2} & = \exp\left (\beta_0^{(2)} + x_{i4}\beta_1^{(2)} + x_{i5}\beta_2^{(2)} +  x_{i6}\beta_3^{(2)}\right)\,.\label{eq:sim:pred2}
\end{align}
Each pair of outcomes $(y_{i1},y_{i2})$ is sampled from a set of different copulae with marginal Poisson parameters $\lambda_{i1}$
 and $\lambda_{i2}$. For each copula, the respective parameter $\theta$ is determined by specifying
 fixed values for Kendall's $\tau\in\{-0.8, -0.6, -0.4, -0.2, 0.1, 0.3, \ldots, 0.9\}$. For \textit{Gumbel} and \textit{Clayton} this can be done analytically via direct transformation of the respective inverse functions, i.e. 
\begin{align*}
\theta_{\text{Gumbel0}} & = \frac{1}{1-\tau}\,, & \tau > 0, \\
\theta_{\text{Clayton0}} & = 2 \frac{\tau}{1-\tau}\,, & \tau > 0, \\
\theta_{\text{Clayton90}} & = -2 \frac{-\tau}{1 + \tau}\,, & \tau < 0,
\end{align*}
while for \textit{Frank}, \textit{Joe} and \textit{Gaussian} this needs to be done numerically solving the following equations:
\begin{align}
\label{tauFrank}
\tau & = 1 - \frac{4}{\theta_{\text{Frank}}} (1 - d_1(\theta_{\text{Frank}}))\,, \\
\label{tauJoe}
\tau & = 1 - 4 \sum_{k=1}^\infty \frac{1}{k(\theta_{\text{Joe0}}\, k + 2) (\theta_{\text{Joe0}}(k-1)+2)}\,, \\
\label{tauGauss}
\tau & = \frac{2}{\pi} \arcsin(\theta_{\text{Gaussian}}).
\end{align}
Term $d_1$ denotes the first Debye function. Formulae (\ref{tauFrank}) and (\ref{tauJoe}) are from \cite{hof:copula:2012} and formula (\ref{tauGauss}) from \cite{lindskog2003kendall}. To each pair of outcomes $(Y_{i1},Y_{i2})$ a bivariate distribution function is assigned from which we sample a single (bivariate) realisation. 

Goodness-of-fit is measured by calculating the mean squared errors for the regression coefficients $\betab$ via
\begin{equation}\label{eq:mse}
 \text{MSE} = \frac{1}{8} \left( \sum_{r=0}^3 \left( \beta_r^{(1)} - \hat{\beta}_r^{(1)}\right)^2 + \left( \beta_r^{(2)} - \hat{\beta}_r^{(2)}\right)^2 \right)\,.
\end{equation}

\subsection{Classical count data set up}\label{sec:sim:basic}

This section shows that the proposed estimation framework with unequal coefficients is able to detect the true copula and estimate the parameters reliably. We define two sets of coefficients, namely ${\betab}^{(1)} =(0.5,0.2,-0.2,0)^{T}$ and
${\betab}^{(2)} =(0.2,-0.3,0.1,0.5)^{T}$, determining the linear predictors in equation~\eqref{eq:sim:pred1} and \eqref{eq:sim:pred2}, respectively,
and a chosen copula from the pre-defined set $\{$C0, C90, F, G0, independence, J0, N$\}$, containing seven different copulae (here C0 denotes the classical Clayton, C90 its 90 degree rotated version, F stands for Frank, G0 for Gumbel, J0 for Joe and N for Gaussian). Similarly to our case study, we fix the sample size to $n=250$ and, as stated in Section~\ref{sec:simulation}, covariates $x_{i1}, \ldots, x_{i6}$ are generated from a uniform $\mathcal{U}[0,1]$ distribution.

Each copula from the aforementioned list is combined with suitable values for Kendall's $\tau$ from the set $\{-0.8, -0.6,\ldots, 0.7, 0.9\}$ and each setting is repeated 100 times. For positive $\tau$, five copulae (N, F, G0, C0, J0) are used, which leads to $100 \times 5 \times 5 = 2500$ samples with $n=250$ bivariate observations each. The three copulae N, F and C90, which can depict negative correlation, are also combined with the four negative $\tau$ values, hence leading to $100 \times 3 \times 4 = 1200$ data sets. The penalisation approach to force equal coefficients in both marginal distributions from Section~\ref{sec:penalty} is not yet applied here. The use of other copulae and respective rotations, implemented in \texttt{GJRM}, led to similar conclusions. 

\begin{figure}[h!]
\includegraphics[width=\textwidth, page=1]{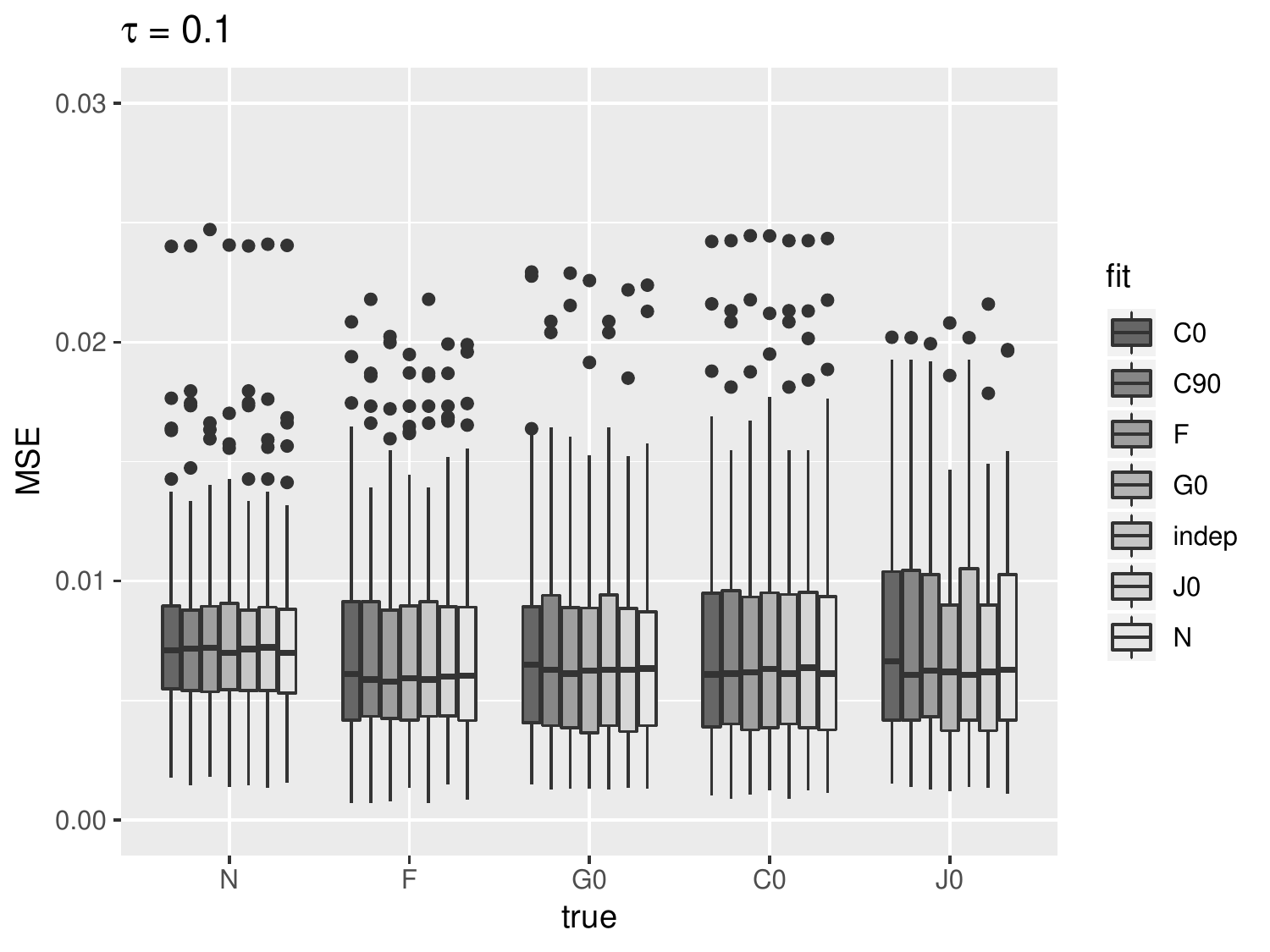}
\caption{Results for MSE of the regression coefficients for different true copulae with each copula parameter $\theta$ derived from $\tau = 0.1$.}
\label{fig:tau:0.1}
\end{figure}

\begin{figure}[h!]
\includegraphics[width=\textwidth, page=4]{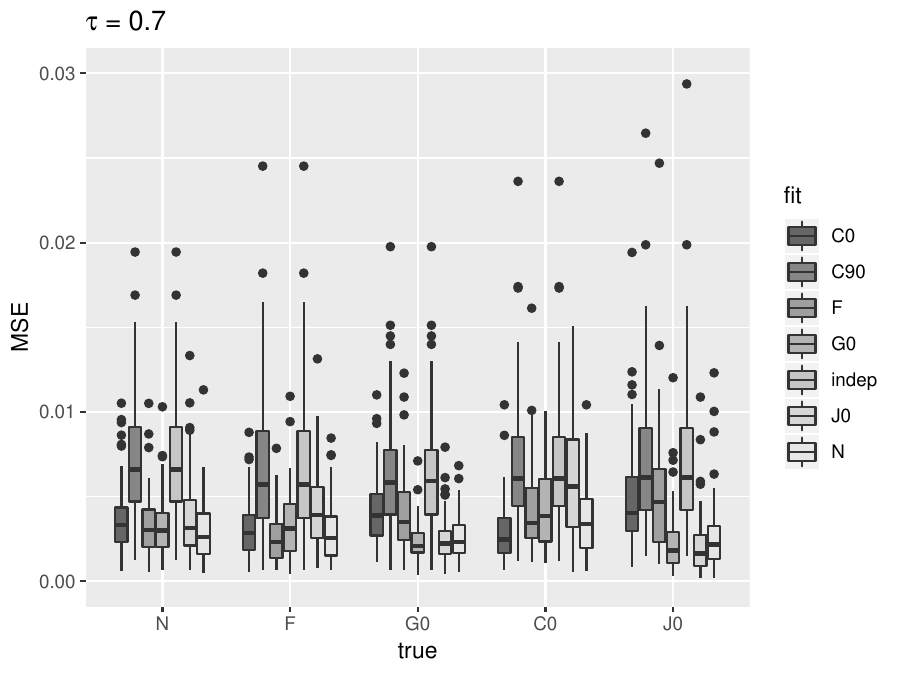}
\caption{Results for MSE of the regression coefficients for different true copulae with each copula parameter $\theta$ derived from $\tau = 0.7$.}
\label{fig:tau:0.7}
\end{figure}

Figure~\ref{fig:tau:0.1} displays boxplots of the resulting MSEs of the regression coefficients from equation~\eqref{eq:mse}
for different true copulae and a selection of fitted ones in a scenario of weak positive correlation ($\tau = 0.1)$.
Due to the weak correlation, the resulting copula structures are similar to an independence
approach and hence no visible differences in goodness-of-fit occur.
When simulating from a setup with a considerably stronger dependence structure, e.g., $\tau = 0.7$, the
results show substantial  differences regarding the selection of the copula function (see Figure~\ref{fig:tau:0.7}).
Stronger association implies a better ability to select the appropriate copula; increasing $\tau$ emphasises each copula's individual shape.

\begin{figure}[h!]
\includegraphics[width=\textwidth, page=8]{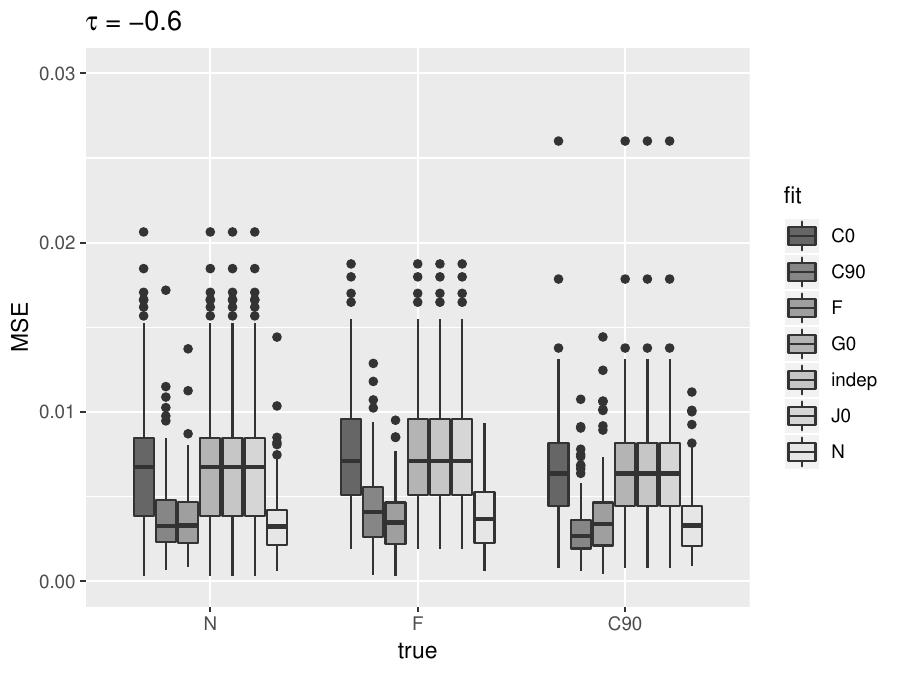}
\caption{Results for MSE of the regression coefficients for different true copulae with each copula parameter $\theta$ derived from $\tau = -0.6$.}
\label{fig:tau:-0.6}
\end{figure}

The results for C90 are numerically identical to those of the independence copula for most samples. Due to the copula's inability to account for positive correlation ($\tau >0$), the fitting process results in $\hat\theta \approx 0$, which practically implies an independence copula. This occurs in both directions; if data are sampled from copulae with $\tau<0$ (see Figure~\ref{fig:tau:-0.6}) the copulae C0, G0 and J0 will lead to fits reflecting (approximately) the independence copula
as they can only account for positive association.

Apart from the direction of the association being an important factor, we found that the proposed approach is able to select the true copula in terms of the regression coefficients' MSEs. Moreover, some copulae lead to results of similar quality. For example, Figure~\ref{fig:tau:0.7} shows that when the data were sampled from a G0 and J0 structure, both copulae deliver satisfying results. In fact, given our setting, these copulae are rather similar in terms of implied dependence structure. Note also that employing the incorrect copula might still lead to good results as long as the general association `direction', be it a positive or negative value of Kendall's $\tau$, can be accounted for. To this end, it is often useful to fit a selection of copulae capturing different types of dependence and then choose a model based on cross-validation or information criteria. Previous research by \citet{fang2014comparison} and \citet{MaRa:2017} suggests that the Akaike's information criterion (AIC; \citealp{Akaike:73}) is able to detect the true underlying copula function. In the following, we investigate the AIC's ability to identify the correct copula structure.

\begin{table}[ht]
\caption{Absolute number of times each copula is chosen by AIC as compared to the true function, for $\tau=0.1$ (maximum in bold).}
\centering
\label{aicresult tau=0.1}
\begin{tabular}{lr|rrrrrrr}
& & \multicolumn{7}{c}{fitted class} \\
& & indep & N & F & G0 &J0 &  C0 & C90  \\
  \hline
 \multirow{5}{*}{true class} & N &  20 &  \textbf{35} &  12 &  13 & 5 &  15 &   0  \\
&  F &  12 &  13 &  \textbf{31} &   7 & 4 &  32 &   1  \\
 & G0 &  10 &  10 &   9 &  \textbf{23} & 43 &   5 &   0  \\
  & J0 &   3 &   4 &   2 &  21 & \textbf{70}&  0 &   0 \\
 & C0 &  20 &   9 &   8 &   6 & 1 & \textbf{55} &   1  \\

   \hline
\end{tabular}
\end{table}

\begin{table}[ht]
\centering
\caption{Absolute number of times each copula is chosen by AIC as compared to the true function, for $\tau=0.7$ (maximum in bold).}
\label{aicresult tau=0.7}
\begin{tabular}{lr|rrrrrrr}
& & \multicolumn{7}{c}{fitted class} \\
&  & indep & N & F & G0 & J0 & C0 & C90 \\
  \hline
 \multirow{5}{*}{true class} & N &   0 &  \textbf{62} &   0 &  27 & 10 &   1 &   0 \\
 & F &   0 &   2 &  \textbf{59} &   0 & 0 &  39 &   0  \\
 & G0 &   0 &  24 &   0 &  \textbf{47} & 29 &  0 &   0  \\
  & J0 &   0 &   6 &   0 &  15 &  \textbf{79} & 0 &   0  \\
 & C0 &   0 &   0 &  13 &   0 & 0 & \textbf{87} &   0  \\
   \hline
\end{tabular}
\end{table}

For a weak dependence structure ($\tau = 0.1$) AIC delivers mixed results (see Table~\ref{aicresult tau=0.1}).
Due to our setting with a small sample size and a rather small range for the response values (the specified covariate distributions and coefficient values yield maximal values for $\lambda_j$ of about $\exp(1.1) \approx 3$ for the Poisson marginals), some of the copula functions will lead to very similar bivariate structures. This is supported by the results displayed in Figure~\ref{fig:tau:0.7}, where the G0 and J0 yield similar results in terms of MSE. Nevertheless, the true copula is always the one that is more frequently selected (see bold numbers on the diagonal). When increasing the strength of dependence to $\tau = 0.7$, Table~\ref{aicresult tau=0.7} shows that copula selection improves, although on a somewhat questionable level. Overall, in 334 out of 500 data sets the AIC was able to detect the true underlying copula. Again, the considered small sample size and a small range for the response values play a role here.

\subsection{Football-like count data}\label{sec:sim:basic}

\begin{figure}[h!]
\includegraphics[width=\textwidth]{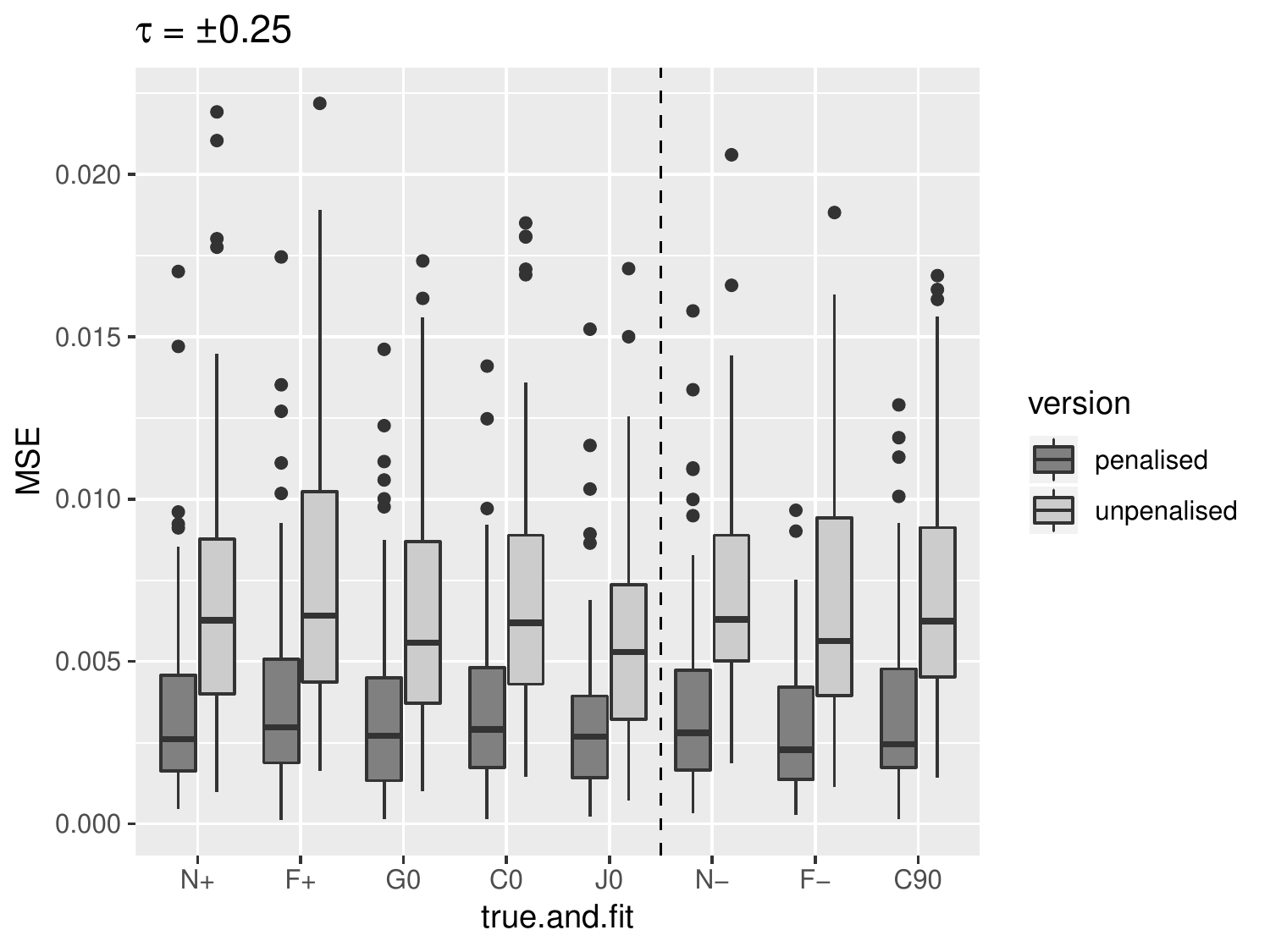}
\caption{Results for the MSE of the regression coefficients obtained using the unpenalised (red) and  penalised (turquois) estimation approaches for a
 set of different copulae and associations; $\tau=0.25$ for copulae N, F, G0, C0, J0 and $\tau = -0.25$ for copulae N, F, C90.}
\label{fig:penalised}
\end{figure}

If both coefficient vectors $\boldsymbol{\beta}^{(1)}$ and $\boldsymbol{\beta}^{(2)}$ are equal or expected to be,
our approach from Section~\ref{sec:penalty} can be used. This is particularly relevant for football (or other sports) data. As compared to the previous simulation set up, coefficients are now chosen as \linebreak ${\betab}^{(1)} = \betab^{(2)} = (0.25,0.2,-0.35,0)^{T}$, hence leading to
\begin{align*}
\lambda_{i1} & = \exp(0.25 + 0.2x_{i1} - 0.35x_{i2} + 0x_{i3})\,, \\
\lambda_{i2} & = \exp(0.25 + 0.2x_{i4} - 0.35x_{i5} + 0x_{i6})\,.
\end{align*}
This setup depicts the same influence of the corresponding covariates in both marginal distributions.
Note that choosing $\beta_{3}^{(j)}=0$ creates a noise variable.
This time, the model is always fitted using the true underlying copula structure only, but
we now distinguish between the unpenalised estimation approach and the penalised approach proposed in Section~\ref{sec:penalty}. The results from 100 simulation runs per copula are visualised in Figure~\ref{fig:penalised}.
They clearly show the superior performance of the penalised approach as compared to the unpenalised one. In the latter, unequal coefficient estimates occur often over the replicates.


\section{FIFA World Cup Application}\label{sec:application}

We will now apply the proposed penalised approach to a real world data set containing FIFA World Cup matches and then compare the method's predictive performance to that of the unpenalised technique (Section~\ref{sec:basic:model}).

\subsection{Data}\label{sec:data}

Model building was based on a data set constructed from the five past FIFA World Cups 2002 -- 2018 with 64 matches each. The basic data set (without the World Cup 2018 data) was introduced and described in detail in \citet{GroSchTut:2015} and \citet{SchauGroll2018}. It was then used in \citet{GroEtAl:WM2018b} to make predictions for the World Cup 2018. We extended the data set by including two more variables, namely \textit{Knockout} and \textit{Titleholder}, which together
with all the other covariates, are described below.

\renewcommand{\labelenumi}{(\alph{enumi})}
\begin{enumerate}

\item \textit{GDP per capita}. This is used as ratio of the GDP per capita for each respective country and the worldwide average GDP per capita (source: \url{https://unstats.un.org/unsd/snaama/Index}).

\pagebreak

\item \textit{Population}. The population size of each country is used as ratio of the global population to take global population growth into account (source: \url{https://data.worldbank.org/indicator/SP.POP.TOTL}).

\item \textit{ODDSET probability}. The odds (taken from the German state betting agency ODDSET) are converted into winning probabilities. Therefore, the variable reflects the probabilities for each team to win the respective World Cup; these odds were calculated before the start of each tournament.

\item \textit{FIFA ranking}. The FIFA ranking provides a ranking system for all national teams measuring the performance of the team over the last four years (source: \url{https://de.fifa.com/fifa-world-ranking/}).

\item \textit{Host}. A dummy variable indicating if a national team is the hosting country.

\item \textit{Continent}. A dummy variable indicating if a national team is from the same continent as the host of the World Cup (including the host itself).

\item \textit{Confederation}. This categorical variable comprises the confederation of the respective team with (in principle) six possible values: Africa (CAF); Asia (AFC); Europe (UEFA); North, Central America and Caribbean (CONCACAF); Oceania (OFC); South America (CONMEBOL). The confederations OFC and AFC had to be merged because in the data set only one team (New Zealand, 2006) from OFC participated in one of the considered World Cups.

\item \textit{(Second) maximum number of teammates}. For each squad, both the maximum and second maximum number of teammates playing together in the same national club.

\item \textit{Average age}. The average age of each squad.

\item \textit{Number of Champions League (Europa League) players}. As a measurement of the success of the players at the club level, the number of players in the semi finals (taking place only a few weeks before the respective World Cup) of the UEFA Champions League (CL) and UEFA Europa League.

\item \textit{Number of players abroad}. For each squad, the number of players playing in clubs abroad (in the season previous to the respective World Cup).

\item \textit{Factors describing the team's coach:} For the coach of each national team, \textit{age} and duration of his \textit{tenure} are observed. Furthermore, a dummy variable is included, whether the coach has the same {\it nationality} as his team or not.

\item \textit{Knockout}. A dummy variable indicating if a match is a knockout one.

\item \textit{Titleholder}. A dummy variable indicating if a team is the current World Championship title holder.

\end{enumerate}

There are, therefore, 18 variables which were collected separately for each WorldCup and each participating team. In our data set each bivariate response $(y_1,y_2)$, representing the number of goals each respective team scored in a certain match, is combined with the respective covariates of both teams. For both teams the same set of covariates is used. A shortened example of the overall data is given in Table \ref{data1}. Our final data set (Table \ref{tab:data:matched}) was created by matching the teams' covariates (Table \ref{tab:covar}) with the match result data (Table \ref{tab:results}).

	\begin{table}[ht!]
\small
\caption{\label{data1} Exemplary table showing the results of four matches (a) and a subset of the covariates of the involved teams (b). The matched data sets for each game are shown in (c).}
\centering
\subfloat[Table of results \label{tab:results}]{
\begin{tabular}{lcr}
  \hline
 &  &  \\
  \hline
FRA \includegraphics[width=0.4cm]{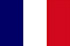} & 0:1 &  \includegraphics[width=0.4cm]{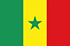} \;SEN\\
URU \includegraphics[width=0.4cm]{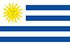} & 1:2 &  \includegraphics[width=0.4cm]{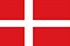} \;DEN\\
DEN \includegraphics[width=0.4cm]{DEN.png} & 1:1 &  \includegraphics[width=0.4cm]{SEN.png} \;SEN\\
FRA \includegraphics[width=0.4cm]{FRA.png} & 0:0 &  \includegraphics[width=0.4cm]{URU.png} \;URU\\

  \vdots & \vdots & \vdots  \\
  \hline
\end{tabular}}
\hspace*{0.8cm}
\subfloat[Table of covariates \label{tab:covar}]{
\begin{tabular}{llrrrrr}
  \hline
World Cup & Team &  Age & Rank & Oddset &   \ldots \\
  \hline
2002 & France  & 28.3 & 1 & 0.149 & \ldots \\
2002 &  Senegal & 24.3 & 42 & 0.006 & \ldots\\
2002 &  Uruguay & 25.3 & 24 & 0.009 & \ldots \\
2002 &  Denmark & 27.4 & 20 & 0.012 & \ldots\\
  \vdots & \vdots & \vdots & \vdots & \vdots  &  $\ddots$ \\
   \hline
\end{tabular}
}\\
\subfloat[Matched data set \label{tab:data:matched}]{
\begin{tabularx}{\textwidth}{ccXXcccrcc}
  \hline
$y_1$ & $y_2$ & Team1 & Team2  & World Cup & Age1 & Age2 & Rank1 & Rank2 &   \ldots \\
  \hline
0 & 1 & FRA \includegraphics[width=0.4cm]{FRA.png}  & SEN \includegraphics[width=0.4cm]{SEN.png}  & 2002  & 28.3 & 24.3 & 1 & 42 & \ldots \\
1 & 2 & URU \includegraphics[width=0.4cm]{URU.png}  & DEN \includegraphics[width=0.4cm]{DEN.png}  & 2002  & 25.3 & 27.4 & 24 & 20 & \ldots\\
1 & 1 & DEN \includegraphics[width=0.4cm]{DEN.png}  & SEN \includegraphics[width=0.4cm]{SEN.png}  & 2002  & 27.4 & 24.3 & 20 & 42 & \ldots \\
0 & 0 & FRA \includegraphics[width=0.4cm]{FRA.png}  & URU \includegraphics[width=0.4cm]{URU.png}  & 2002  & 28.3 & 25.3 & 1 & 24 & \ldots\\

  \vdots & \vdots & \vdots & \vdots &   \vdots & \vdots & \vdots & \vdots & \vdots  &  $\ddots$ \\
  \hline
\end{tabularx}
}
\end{table}

Next, we will fit the unpenalised and penalised versions of the proposed estimation method to the FIFA World Cup data from Table \ref{tab:data:matched}, and use a
cross-validation-type strategy to compare their performance.

\subsection{Comparison of predictive performance}\label{sec:pred:perf}

With prediction ability being our main objective, we have to validate all possible models with
out-of-sample data. To do so, we fit the models (with different copulae) on four out of five World Cups and predict the marginal parameters $\lambda_1,\lambda_2$ for the matches of the left-out tournament in a cross-validation-type strategy, cycling through all five tournaments. Using the estimate for $\theta$ of each copula, we can obtain probabilities for each possible match result $(y_1,y_2)$ via equation~\eqref{eq:copula:pred} from Section~\ref{sec:predict}.

Probabilities for the three-way results \textit{win}, \textit{draw} and \textit{loss} are computed by aggregating all corresponding results; for example, for a draw we sum up the probabilities for the results $(0,0), (1,1), \ldots, (20,20)$, cutting of at a reasonable maximum number of goals. We settled at 20 as cut-off, due to the maximal estimates of $\hat\lambda_j\approx 3$. The same strategy is applied for all the match results that lead to a win or a loss. Note that our estimated three-way probabilities practically always added up to one, which indicates that limiting the analysis to all results up to 20 goals was sufficient. The resulting {\it three-way probabilities} are denoted as $\hat{\pi}_{i l}$ for the $i$-th match and $l = 1,2,3$ indicates \textit{win}, \textit{draw} and \textit{loss}. Such three-way-outcomes could also have been modelled directly by using models for categorical responses. 
However, our approach to model the exact number of goals exploits more information, which is why ordinal/multinomial models were not considered.

For the estimation approaches considered, we employed several measures to compare their predictive performance. Similar to \citet{SchauGroll2018}, we use the rank probability score (RPS), the multinomial likelihood and the classification rate as goodness-of-prediction measures, and average the results over all matches of all cross-validation cycles. In our setting, for a single game the RPS is defined via
\[
\text{RPS}_i \ = \ \frac{1}{2} \sum_{r=1}^2 \left( \sum_{l = 1}^r \hat{\pi}_{i l} - \delta_{i l}\right)^2\,.
\]
The true result is a dummy coding for \textit{win}, \textit{draw}, \textit{loss} and denoted by Kronecker's delta $\delta_{l i}$. The mutinomial likelihood for a single match is defined as
\[
\text{LLH}_i = \hat{\pi}_{i1}^{\delta_{i1}} \hat{\pi}_{i2}^{\delta_{i2}} \hat{\pi}_{i3}^{\delta_{i3}}\,,
\]
which is essentially the predicted probability $\hat{\pi}_{i l}$ for the true outcome. Additionally, the classification rate is given as
$$
\text{CR}_{i} = \mathbb{I}(\tilde y_i=\underset{l\in\{1,2,3\}}{\mbox{arg\,max }}(\hat\pi_{il}))\,,
$$ indicating whether match $i$ was correctly classified. All measures are calculated for the unpenalised and penalised models and for each match prediction, and are then averaged over all $n=320$ FIFA World Cup matches.


Beside focusing on three-way-outcomes, we can also analyse the performance in predicting the exact number of goals each team scored. Hence, we include 
the Euclidean distance between observation and prediction. With the bivariate mean $\lambdab_i = (\lambda_{i1}, \lambda_{i2})^T$ of the bivariate distribution for a single match corresponding to a given copula model, we have 
\[ 
\text{MSE} = \frac{1}{n}\sum_{i=1}^n \norm{ \yb_i - \hat{\lambdab}_i}_2 \,\,\, =  \,\,\, \frac{1}{n} \sum_{i=1}^n \sqrt{(y_{i1}-\hat{\lambda}_{i1})^2 + (y_{i2}-\hat{\lambda}_{i2})^2},
\]
which is indeed calculated over $n$ observed and predicted match results.


As predictive ability is our main aim, we also investigate the betting outcome for the most recent FIFA World Cup 2018 as another measure of performance.
Using the (average) betting odds of the 64 matches (obtained from \url{oddsportal.com}, which provides averaged three-way-odds from a selection of bookmaker companies) as well as the corresponding predicted probabilities from our respective models, different betting strategies can be applied (see, e.g., \citealp{GrollEtAl2018}). For every match $i$ and each of the possible three outcomes $l\in\{1,2,3\}$, one can
calculate the expected return as follows: $E[return_{il}]=\hat \pi_{il}*odds_{il}-1$. Then, one could choose the outcome with the highest expected return, but only place the bet if the expected maximum return is positive, i.e.\ if $\max\limits_{l\in\{1,2,3\}}E[return_{il}] > \varepsilon$ holds, with $\varepsilon=0$. \citet{koop:2015}
 used different values of the threshold $\varepsilon>0$ and showed that this way the overall average return could
 be increased. While they used constant stake sizes (one arbitrary unit) for each bet, alternative
betting strategies with varying stake sizes based on Kelly's criterion \citep{Kelly:1956} can applied; see for example \citet{boshnakov2017}.
With this criterion the optimal stake for single bets can be determined in order to maximize the
return by considering the size of the odds and the winning probability.
We will use the simple strategy with constant betting stakes and $\tau = 0$ in our copula selection process
and provide a more detailed look into results afterwards.

\subsection{Results}\label{sec:appl:results}

\begin{figure}[t!]
\centering
\caption{Histograms and normal Q–Q plots of randomised normalized quantile residuals for the margins produced after fitting a Gaussian
copula model with Poisson-distributed marginals to data from all World Cups.}
\label{qqplots}
\includegraphics[scale=0.7]{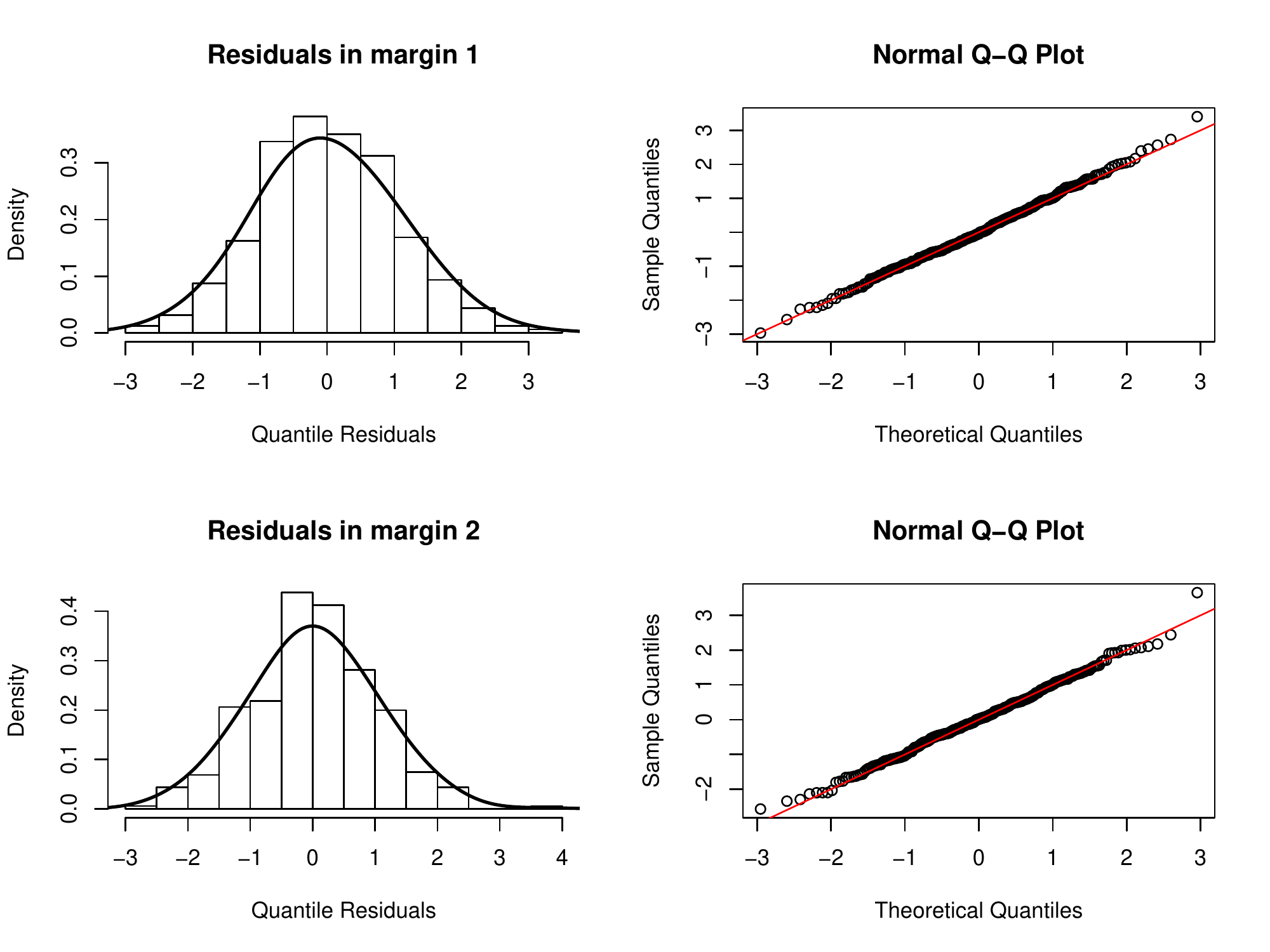}
\end{figure}

The assumption of Poisson-distributed margins can be checked using randomised normalised quantile residuals \citep[see][and references therein]{MaRa:2017}. In this case, using for instance a Gaussian copula model with Poisson margins fitted to data from all World Cups, the choice of marginal distributions seems appropriate (see Figure \ref{qqplots}).

The results for all the performance measures described in the previous section can be found in Table~\ref{resultsfootball} in Appendix A. For each copula, the predictive performance of the penalised estimation approach is substantially better than that of the unpenalised method.

In our setup, the selected measures yield their bests values (highlighted in bold font) for different copulae, namely T (Student's T copula with 3 degrees of freedom), N (Gaussian copula), J90 (Joe copula rotated by 90 degrees), and even an independence model fitted outside of the \texttt{GJRM} framework. To aid comparability one could aggregate for each copula the corresponding ranks of all the considered measures (taking the different layouts of the selected measures into account). The respective results can be found in Table \ref{rank:result} where we see that F (Frank copula), followed by FGM (Farlie-Gumbel-Morgenstern copula) and N are the best for this specific scenario.

\begin{table}[b!]
\centering
\caption{Ranks according to selected measures for fits based on different copulae.}
\label{rank:result}
\begin{tabularx}{\textwidth}{|Xrrrrrr||Xrrrrrr|}
  \hline
\rowcolor{gray!50} Cop & RPS & LLH & CR & bets & MSE & $\Sigma$ & Cop & RPS & LLH & CR & bets & MSE & $\Sigma$ \\
  \hline
F &   6 &   3 &  14 &   5 &  10 &  38 & G180 &   7 &   5 &  12 &   5 &  17 &  46 \\ 
  FGM &   8 &   6 &  14 &   2 &   8 &  38 & J180 &   5 &   8 &  12 &   5 &  18 &  48 \\ 
  N &   9 &   9 &  10 &   1 &  12 &  41 & J270 &  17 &  17 &   2 &   9 &   4 &  49 \\ 
  J90 &  15 &  15 &   2 &   9 &   1 &  42 & G0 &  10 &  10 &  11 &   4 &  15 &  50 \\ 
  G270 &  14 &  14 &   2 &   9 &   3 &  42 & AMH &   2 &   4 &  14 &  17 &  14 &  51 \\ 
  C180 &  11 &  11 &   9 &   3 &   9 &  43 & C270 &  18 &  18 &   2 &   9 &   5 &  52 \\ 
  indep &  13 &  13 &   1 &   9 &   7 &  43 & PL &   4 &   2 &  18 &  17 &  13 &  54 \\ 
  C0 &   3 &   7 &  14 &   5 &  16 &  45 & C90 &  19 &  19 &   2 &   9 &   6 &  55 \\ 
  G90 &  16 &  16 &   2 &   9 &   2 &  45 & T &   1 &   1 &  19 &  17 &  19 &  57 \\ 
  J0 &  12 &  12 &   2 &   9 &  11 &  46 &  &  &  &  &  &  &  \\ 
   \hline
\end{tabularx}
\end{table}

Compared to the results of the models with no dependence structure (independence copula model obtained via a univariate Poisson regression approach on the single numbers of goals with three-way probabilities estimated via the Skellam distribution), we can see that the copula models improve the values for the chosen measures by a small margin or not at all. This was expected since a relatively weak dependence structure in the scores of international football matches
was shown in previous work (see, e.g., \citealp{GrollEtAl2018} who found that no additional dependence modelling was needed in a bivariate Poisson model when suitably structured predictors are employed). The estimated parameter $\hat{\theta}=0.904$ and its corresponding value of Kendall's $\hat\tau = 0.100$ indicate a rather weak dependence structure for the F copula. The second and third ranked copula models with estimated values of $\hat{\theta}=0.405$ leading to $\hat\tau = 0.09$ for FGM and $\hat{\theta}=0.116$ leading to $\hat\tau = 0.0738$ for N (fitted on all World Cups) support the presence of a rather weak dependence structure. Table~\ref{tab:coefs} in Appendix~\ref{App:Tables} shows the estimated regression coefficients for the F copula model.

Using the AIC for copula selection did not confirm the previous results: PL (Plackett copula) -- which achieved the 17th place with respect to our five prediction measures (RPS, likelihood, classification, betting results, MSE) -- provided the best fit. The F and FGM copulae, however, are ranked 3rd and 7th according to the AIC, and performed the best among our measures. It is important to stress that when using information criteria (but not only) the selection of the copula function is expected to improve as the sample grows large. Because predictive ability is our main goal, we rely more on the aforementioned measures over the AIC. The next paragraph shows that it can be advantageous to use the proposed copula models for betting strategies.

\medskip

\medskip

\begin{figure}[h!]
\begin{center}
\includegraphics[width=0.8\textwidth]{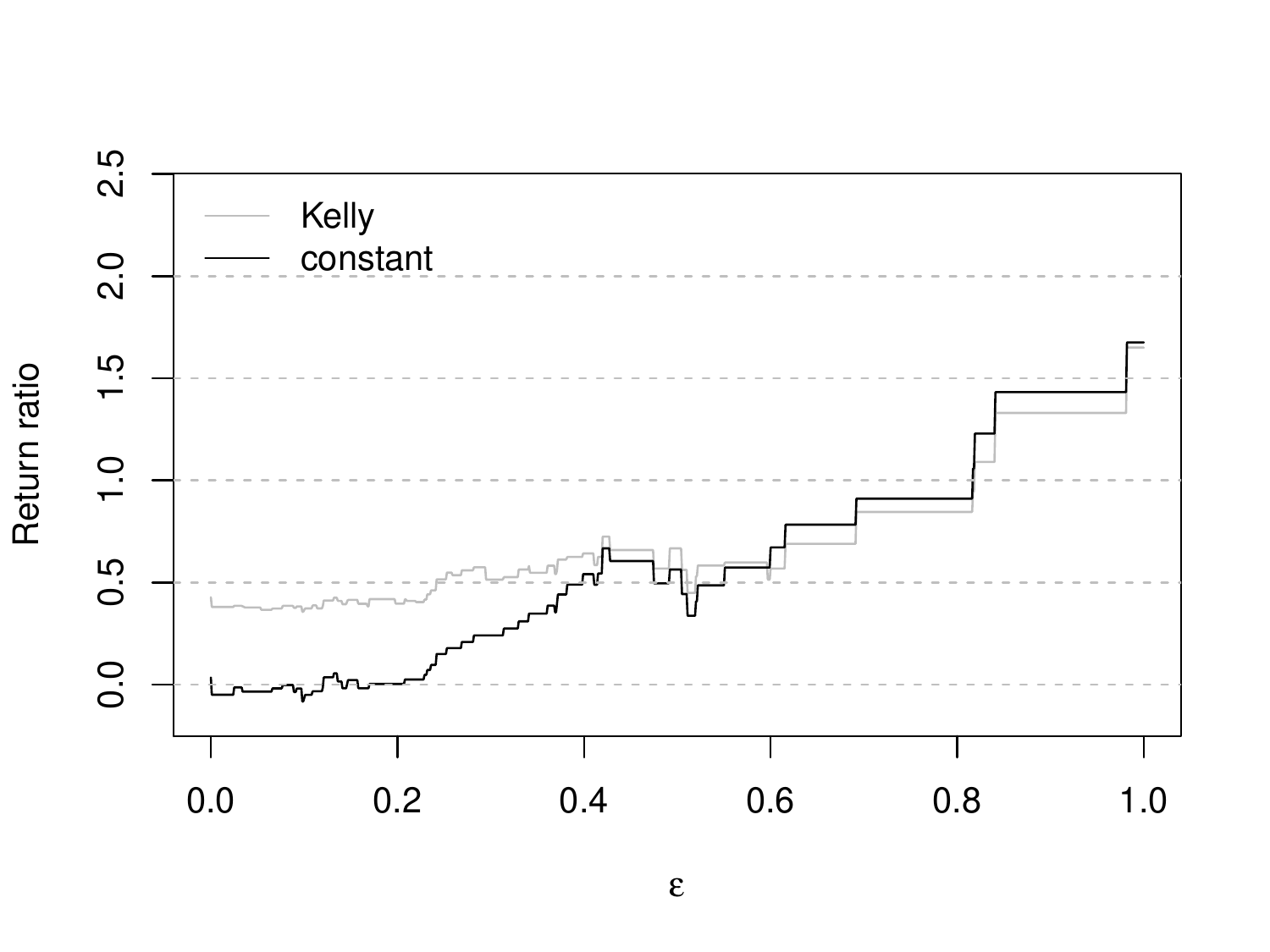}\vspace{-2cm}
\includegraphics[width=0.8\textwidth]{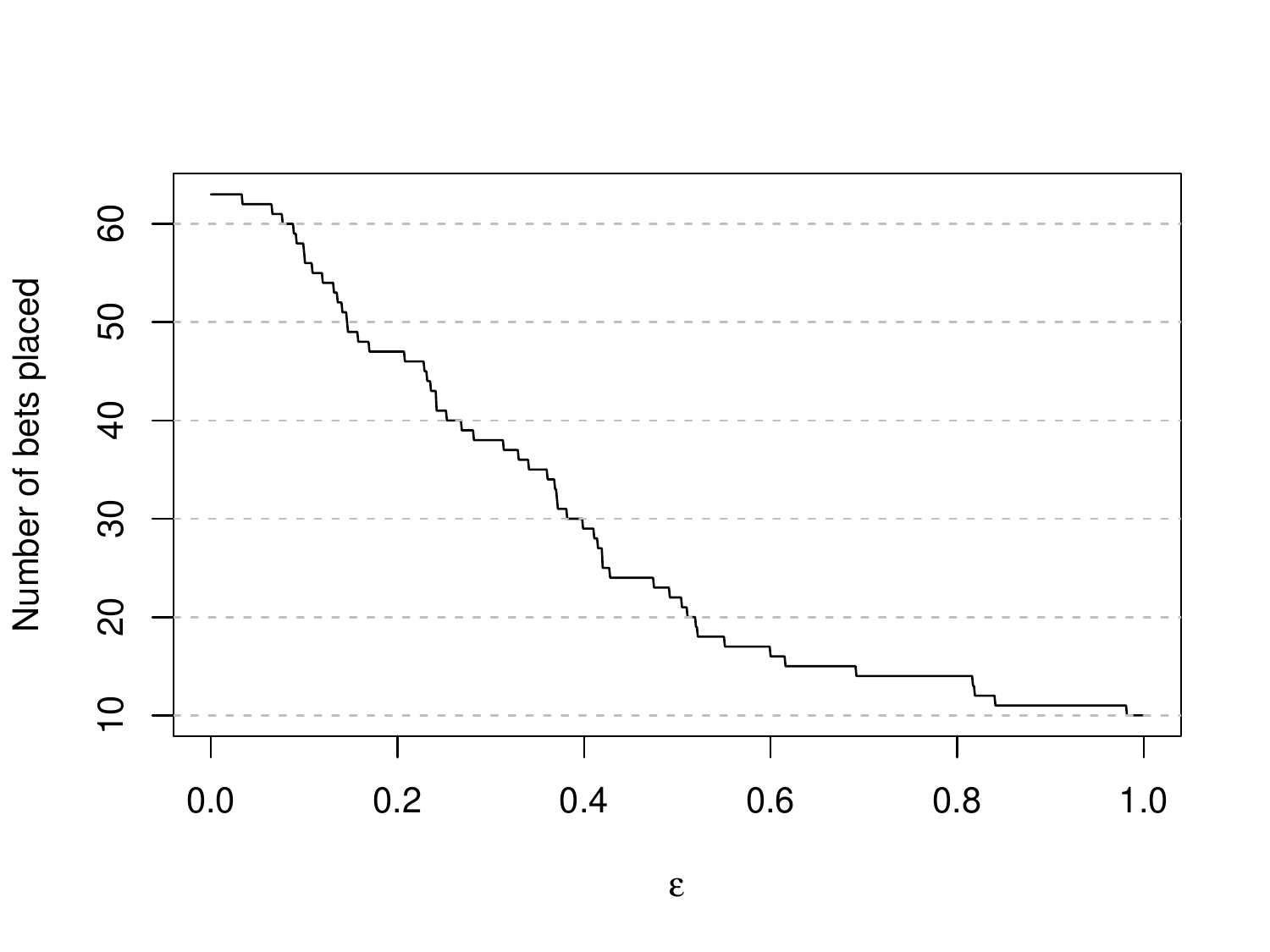}\vspace{-0.5cm}
\caption{Top: Return ratios for different betting strategies vs.\ threshold $\varepsilon$;
Bottom: Number of bets placed vs.\ threshold  $\varepsilon$}
\label{fig:betting}
\end{center}
\end{figure}

\noindent{\bf Betting}\\
Fictional betting results can be calculated by predicting the World Cup 2018 outcomes from the F copula model fitted on World Cups 2002 -- 2014. Figure \ref{fig:betting} (top) depicts the average return percentage (i.e., the ratio between profit and
investment) of two strategies for varying threshold sizes $\varepsilon\geq 0$. Note that with increasing values of $\varepsilon$
the number of matches on which bets are placed decreases (see bottom of Figure \ref{fig:betting}). For the FIFA World Cup 2018, solid positive returns can be achieved for values of $\varepsilon > 0.25$ with a simple betting strategy of constant stakes. Expanding this strategy with flexible stakes via Kelly's approach leads to positive returns for all $\varepsilon$. Overall, Kelly's strategy is clearly superior to placing constant stakes indicated by higher (or equal) returns especially for smaller values of $\varepsilon$. An investment of 100 units of an arbitrary currency with a betting strategy of a fixed $\varepsilon = 0.4$ would yield a profit of about
50 units with constant stakes and 60 units with flexible stakes via Kelly's approach. Though remarkable at a first glance, these results have to be analysed cautiously. Due to the rather small sample size, the betting results very much depend on single match results and are probably very variable, especially for higher values of $\varepsilon$ and therefore a smaller number of placed bets. Note, in fact, that the model is prone to extreme betting odds. For example, the bookmaker's odds for South Korea's victory against Germany were on average 19.52. Our model would recommend to bet on such outcomes - thus the betting results are likely to suffer from high variability. Despite these limitations, the results of the betting strategies are in favour of copula-structured models.

\section{Conclusion}\label{sec:conclusion}

In this work, we have proposed a generalised joint regression framework for count responses.
The method allows for linear and non-linear dependence structures through the use of different copulae, and for each parameter of the model to be specified as function of flexible covariate effects. We have also provided an extension which forces the regression coefficients of the marginal
(linear) predictors to be equal via the use of a suitable penalisation. This is relevant for modelling team sports data (e.g., from football or handball), because otherwise being first- or second-named team could affect the regression coefficients' estimates and their interpretation. The proposed method is available via the \texttt{R} add-on package \texttt{GJRM}.

We investigated the method's performance in two simulation studies, the first one designed for arbitrary count data, the other one reflecting football-specific settings. In the first study, when the outcomes are weakly associated, copula structures are similar to an independence model and hence no tangible differences in goodness-of-fit are observed. With stronger dependence between the outcomes, results show substantial gains when using copula models. Generally, we found that the proposed method is able to select the true copula in terms of evaluating the regression coefficients' MSE. In the second simulation study, we assumed equal coefficients for the two marginal distributions. Under this scenario, the penalised method delivers an improved performance as compared to the unpenalised technique.

The method was applied to FIFA World Cup data; by using a cross-validation-type strategy based on several prediction measures, the penalised version of the method clearly outperformed the unpenalised approach. Moreover, the penalised approach performed better with regard to certain betting strategies.

Future research will address several extensions. Firstly, although not yet considered in our case study, 
we would like to extend the penalty discussed in this paper to the context of more complex predictor structures (allowing, for instance, for non-linear effects via P-splines). Moreover, we believe that the method's predictive performance can be further improved by penalising covariate effects via LASSO-type penalties \citep{Tibshirani:96, FrieEtAl:2010} or via boosting \citep[e.g.,][]{BueHo:2007, HotEtAl:2010}, a technique that stems from machine learning. These methods already proved to be effective in the context of predicting football matches \citep[e.g.,][]{GroSchTut:2015, GrollEtAl2018}.

\bibliographystyle{chicago}
\bibliography{literatur}
\newpage
\appendix
\section*{Appendix}
\section{Software}
\label{software}
Copula models with discrete margins (but not only) can be fitted using the \texttt{GJRM} package by \citet{Mar:2017:gjrm} for the \texttt{R} environment (\citealp{RDev:2019}). The syntax is similar to that of established methods and packages for generalised linear and additive models. In the following, we provide an example. The model's formulae are provided, for instance, as
\begin{verbatim}
eq1 <- Goals ~ CL.players + UEFA.players + Nation.Coach + Age.Coach + 
	  Tenure.Coach + Legionaires + max.teammates + sec.max.teammates + age + 
	  Rank + GDP + host + confed + continent + odds + Population + Knockout + 
	  titleholder
eq2 <- Goals.oppo ~ CL.players.oppo + UEFA.players.oppo + 
	  Nation.Coach.oppo + Age.Coach.oppo + Tenure.Coach.oppo + 
	  Legionaires.oppo + max.teammates.oppo + sec.max.teammates.oppo +
	  age.oppo + Rank.oppo + GDP.oppo + host.oppo + confed.oppo + 
	  continent.oppo + odds.oppo + Population.oppo + Knockout + 
	  titleholder.oppo
eq3 <- ~ 1
eqlist <- list(eq1, eq2, eq3)
\end{verbatim}
where \texttt{Goals, Goals.oppo} are the discrete responses and the variables on the right hand side represent the covariates (regressors can either have or not the suffix \texttt{oppo}, depending on the margin considered). The same covariate can be used in more than one equation if desired (e.g., \texttt{Knockout}). The model is fitted by the call
\begin{verbatim}
fit <- gjrm(eqlist, data = WorldCup, BivD = "N", Model = "B", 
	  margins = c("PO", "PO"))
\end{verbatim}
with \texttt{BivD} denotes the chosen copula (here, \texttt{N} for Gaussian), and the \texttt{margins} are Poisson distributed. Flexible covariate effects can be accounted for via splines by using for example something like \texttt{s(covariate)}. 

Convenience functions such as \texttt{summary()} and \texttt{plot()} are used in the same fashion as those for generalised linear and additive models. Residual diagnostics such as those displayed in Figure \ref{qqplots} can be obtained via \texttt{post.check()}.

More details, options, and extensive examples are given in the documentation of the \texttt{GJRM} package.

For this work the function \texttt{gjrm()} was adapted to allow for a new boolean argument \texttt{linear.equal}. For future work flexible penalisation utilities (including the proposed one in this paper) are planned to be implemented into the \texttt{GJRM} package.

\newpage
\section{Tables}
\label{App:Tables}

\rowcolors{2}{gray!15}{white}
\begin{table}[h!]
\centering
\caption{Results of selected measures for model fits based on different copulae obtained using the unpenalised and penalised approaches.}
\label{resultsfootball}
\begin{tabularx}{\textwidth}{|X|XXXXXXrrXX|}
  \hline

\rowcolor{gray!50}   Cop-    & \multicolumn{2}{c}{RPS}   & \multicolumn{2}{c}{likelihood} & \multicolumn{2}{c}{class. rate} & \multicolumn{2}{c}{betting} & \multicolumn{2}{c|}{$\text{MSE}_{\text{Goals}}$} \\
\rowcolor{gray!50} ula  & pen & unp & pen & unp & pen & unp & pen & unp & pen & unp \\
  \hline
N & 0.196 & 0.210 & 0.403 & 0.395 & 0.522 & 0.506 & \textbf{0.199} & -0.225 & 1.421 & 1.490 \\ 
  C0 & 0.196 & 0.210 & 0.404 & 0.396 & 0.512 & 0.484 & 0.035 & -0.154 & 1.424 & 1.496 \\ 
  C90 & 0.198 & 0.211 & 0.398 & 0.390 & 0.528 & 0.509 & -0.012 & -0.240 & 1.418 & 1.486 \\ 
  C180 & 0.198 & 0.212 & 0.400 & 0.392 & 0.525 & 0.506 & 0.040 & -0.240 & 1.421 & 1.490 \\ 
  C270 & 0.198 & 0.212 & 0.398 & 0.390 & 0.528 & 0.506 & -0.012 & -0.240 & 1.418 & 1.486 \\ 
  J0 & 0.198 & 0.211 & 0.400 & 0.392 & 0.528 & 0.500 & -0.012 & -0.240 & 1.421 & 1.490 \\ 
  J90 & 0.198 & 0.212 & 0.398 & 0.390 & 0.528 & 0.506 & -0.012 & -0.240 & \textbf{1.418} & 1.486 \\ 
  J180 & 0.196 & 0.210 & 0.404 & 0.396 & 0.516 & 0.478 & 0.035 & -0.225 & 1.425 & 1.500 \\ 
  J270 & 0.198 & 0.211 & 0.398 & 0.390 & 0.528 & 0.509 & -0.012 & -0.240 & 1.418 & 1.486 \\ 
  G0 & 0.197 & 0.212 & 0.402 & 0.394 & 0.519 & 0.503 & 0.038 & -0.278 & 1.422 & 1.492 \\ 
  G90 & 0.198 & 0.211 & 0.398 & 0.390 & 0.528 & 0.509 & -0.012 & -0.240 & 1.418 & 1.486 \\ 
  G180 & 0.196 & 0.210 & 0.404 & 0.396 & 0.516 & 0.484 & 0.035 & -0.236 & 1.424 & 1.495 \\ 
  G270 & 0.198 & 0.211 & 0.398 & 0.390 & 0.528 & 0.509 & -0.012 & -0.240 & 1.418 & 1.486 \\ 
  F & 0.196 & 0.210 & 0.405 & 0.396 & 0.512 & 0.494 & 0.035 & -0.284 & 1.421 & 1.487 \\ 
  AMH & 0.196 & 0.210 & 0.405 & 0.396 & 0.512 & 0.491 & -0.049 & -0.110 & 1.421 & 1.489 \\ 
  FGM & 0.196 & 0.210 & 0.404 & 0.396 & 0.512 & 0.491 & 0.117 & -0.223 & 1.420 & 1.486 \\ 
  T & \textbf{0.195} & 0.212 & \textbf{0.407} & 0.398 & 0.506 & 0.469 & -0.049 & -0.219 & 1.430 & 1.500 \\ 
  PL & 0.196 & 0.210 & 0.405 & 0.396 & 0.509 & 0.484 & -0.049 & -0.284 & 1.421 & 1.487 \\ 
  indep & 0.198 & 0.211 & 0.398 & 0.390 & \textbf{0.531} & 0.509 & -0.012 & -0.240 & 1.419 & 1.486 \\
   \hline
\end{tabularx}
\end{table}

\begin{table}[h!]
\centering
\caption{Estimated coefficients and standard errors for the FGM model fitted on all World Cups.}
\label{tab:coefs}
\begin{tabularx}{\textwidth}{|Xrr||Xrr|}
  \hline
  \rowcolor{gray!50} Covariate & \multicolumn{1}{c}{$\hat{\beta}$} & se$(\hat{\beta})$ & Covariate & \multicolumn{1}{c}{$\hat{\beta}$} & se$(\hat{\beta})$\\
  	Intercept & 2.226 & 1.078 & GDP & 0.042 & 0.026  \\
	CL Players & 0.038 & 0.025  & Host & 0.369 & 0.186  \\
	EL Players & 0.048 & 0.027 & conf. CAF & 0.094 & 0.206  \\
	Nationality Coach & 0.097 & 0.098 & conf. CONCACAF & 0.109 & 0.211  \\
	Age Coach & -0.008 & 0.005 & conf. CONMEBOL & 0.622 & 0.217 \\
	Tenure Coach & -0.043 & 0.022 & conf. UEFA & 0.395 & 0.184 \\
	Players Abroad & 0.005 & 0.011 & Continent & -0.023 & 0.088 \\
	Max. Teammates & -0.011 & 0.043 & Odds & -0.593 & 1.521  \\
	2nd Max. Teammates & 0.004 & 0.064 & Population & 0.069 & 0.042\\
	Age & -0.054 & 0.038 & Knockout & -0.443 & 0.096 \\
	Rank & -0.007 & 0.004 & Titleholder & -0.274 & 0.262 \\
   \hline
\end{tabularx}
\end{table}

\end{document}